\documentclass[prl,unsortedaddress,superscriptaddress]{revtex4}
\usepackage{amsmath}  
\usepackage{amsbsy,amssymb,amsfonts}
\usepackage{graphicx}
\usepackage{color}
\usepackage{epsf}
\usepackage[utf8]{inputenc}
\usepackage{hyperref}
\usepackage{multirow}
\def\ie{{\em{i.e.}},}

\def\cm{$\rm cm^{-1}$}

\def\bravert{\egroup\,\vrule\,\bgroup}

{\catcode`\|=\active
  \gdef\Twoint#1{\left(\mathcode`\|"8000\let|\bravert {#1}\right)}}
{\catcode`\|=\active
  \gdef\Braket#1{\left<\mathcode`\|"8000\let|\bravert {#1}\right>}}

\newcommand{\beq}{\begin{equation}}
\newcommand{\eeq}{\end{equation}}
\newcommand{\beqa}{\begin{eqnarray}}
\newcommand{\eeqa}{\end{eqnarray}}
\newcommand{\bea}{\begin{array}}
\newcommand{\eea}{\end{array}}

\newcommand{\bef}{\begin{figure}}
\newcommand{\ef}{\end{figure}}
\newcommand{\bc}{\begin{center}}
\newcommand{\ec}{\end{center}}
\newcommand{\bt}{\begin{table}}
\newcommand{\et}{\end{table}}
\newcommand{\btb}{\begin{tabular}}
\newcommand{\etb}{\end{tabular}}

\def\rvac{\left| \rule{0.3cm}{.0cm} \right>}
\def\lvac{\left< \rule{0.3cm}{.0cm} \right|}
\def\molcas{$\cal M\kern-0.10em O\kern-0.15em L\kern-0.00em 
             C\kern-0.10em A\kern-0.05em S$}
\begin{document}

\vspace{2cm}
\title {{
         TaN, a molecular system for probing \\ ${\cal{P,T}}$-violating hadron physics
       }}

\vspace*{2cm}

\author{Timo Fleig}
\email{timo.fleig@irsamc.ups-tlse.fr}
\affiliation{Laboratoire de Chimie et Physique Quantiques,
             IRSAMC, Universit{\'e} Paul Sabatier Toulouse III,
             118 Route de Narbonne, 
             F-31062 Toulouse, France }

\author{Malaya K. Nayak}
\email{mk.nayak72@gmail.com}
\affiliation{Bhabha Atomic Research Centre,
             Trombay, Mumbai - 400085,
             India}
\affiliation{Laboratoire de Chimie et Physique Quantiques,
             IRSAMC, Universit{\'e} Paul Sabatier Toulouse III,
             118 Route de Narbonne, 
             F-31062 Toulouse, France }

\author{Mikhail Kozlov}
\email{mgk@mf1309.spb.edu}
\affiliation{Petersburg Nuclear Physics Institute, Gatchina 188300, Russia}
\affiliation{St. Petersburg Electrotechnical University ``LETI'', Professor Popov Street 5,
             St. Petersburg 197376, Russia}

\date{\today}
\vspace*{1cm}
\begin{abstract}
All-electron configuration interaction theory in the framework of the Dirac-Coulomb Hamiltonian
has been applied
to the TaN molecule, a promising candidate in the search for Beyond-Standard-Model physics
in both the hadron and the lepton sector of matter. We obtain in the first excited {$^3\Delta_1$} 
state a ${\cal{P,T}}$-odd effective electric field of $36.0 \left[\frac{\rm GV}{\rm cm}\right]$,
a scalar-pseudoscalar ${\cal{P,T}}$-odd interaction constant of $32.8$ [kHz], and a nuclear
magnetic-quadrupole moment interaction constant of $0.74$ [$\frac{10^{33} {\text{Hz}}}{e\, {\text{cm}}^2}$].
The latter interaction constant has been obtained with a new approach which we describe in detail.
Using the same highly correlated all-electron wavefunctions with up to $2.5$ billion expansion
terms, we obtain a parallel magnetic
hyperfine interaction constant of $-2954$ [MHz] for the $\rm {^{181}{T}a}$ nucleus, a very
large molecule-frame electric dipole moment of $-4.91$ [Debye], and spectroscopic constants for
the four lowest-lying electronic states of the molecule.

\end{abstract}

\maketitle
\clearpage

\section{Introduction}
\label{SEC:INTRO}
Electric Dipole Moments (EDMs) are sensitive low-energy probes for New Physics, i.e., physics 
beyond the Standard Model (SM) of elementary particles \cite{ramsey-musolf_review2_2013}.
The existence of an EDM gives rise to ${\cal{P}}$- (spatial parity) and ${\cal{T}}$- 
(time-reversal) odd interactions \cite{khriplovich_lamoreaux} which originate in fundamental 
sources of (${\cal{CP}}$)-violation \cite{EDMsNP_PospelovRitz2005} and are related to these via the 
${\cal{CPT}}$-theorem \cite{streater}.

Atomic and molecular systems allow for probing (${\cal{CP}}$)-violation both in the lepton as
well as the hadron sector of matter. In the former case, typically the electron EDM and the
electron-nucleon scalar-pseudoscalar (eN-SPS) interaction are investigated 
\cite{commins_demille_EDM_2008}. These two effects are in general the dominant expected 
manifestations of ${\cal{P,T}}$-violation in paramagnetic systems 
\cite{ACME_ThO_eEDM_science2014,hudson_hinds_YbF2011,regan_demille_2002}. Concerning the latter, 
possible important sources of symmetry-violating effects at the atomic scale are
the nuclear Schiff moment \cite{swallows_Hg_PRA2013,khriplovich_lamoreaux} and the nuclear
magnetic quadrupole moment (MQM) \cite{Sushkov_Flambaum_Khriplovich1984}, for the same underlying 
sources of (${\cal{CP}}$)-violation. The EDM due to a nuclear-MQM - electronic-magnetic-field
interaction can for certain types of atomic and molecular systems even be larger than that due to 
the Schiff moment.

In the present paper we investigate the neutral diatomic molecule TaN which has been identified
as a promising candidate in EDM searches \cite{Flambaum_DeMille_Kozlov2014} for various reasons.
Deformed nuclei such as the {$^{181}$Ta} nucleus exhibit a strong enhancement of the MQM relative 
to that of spherical nuclei, roughly one order of magnitude, which increases the prospect of measuring an 
ensuing ${\cal{P,T}}$-odd shift. The MQM is directly related to the quantum chromodynamics 
(QCD) (${\cal{CP}}$)-violating parameter $\tilde{\Theta}$ \cite{Veneziano_Witten_QCD-PTodd1980}.
Furthermore, an MQM can arise {\it{via}} EDMs of the $u$ and $d$ quarks, $d_{u,d}$, and {\it{via}}
chromo-EDMs, $\tilde{d}_{u,d}$ \cite{Gunion_Wyler_CEDM_NEDM_1990} in the framework of spontaneous 
${\cal{CP}}$-violation in the Higgs sector \cite{khriplovich_lamoreaux}. The search for the ensuing
${\cal{P,T}}$-violating effects in a molecular system may, therefore, further constrain (or unravel) 
beyond-standard-model (BSM) hadron physics. A further asset is the electronic structure of the TaN 
molecule. From earlier experimental and theoretical studies it is known that TaN has an energetically 
very low-lying {$^3\Delta$} state \cite{TaN_Bernath_2002} arising mainly from a $\sigma^1\, \delta^1$
electronic configuration. In other valence-isoelectronic candidate molecules such as ThO 
\cite{demille_ThO_PRA2011,Fleig2014,Skripnikov_ThO_JCP2013} and 
HfF$^+$ \cite{Cossel_HfF+_CPL2012,fleig_nayak_eEDM2013} this state is deeply bound and exhibits a
very small magnetic moment which facilitates experimental studies and helps reduce the vulnerability 
of experiments to decoherence and systematic errors \cite{Leanhardt_EDM_JMS2011,Meyer_Bohn_PRA2008}.

In the present work we have pursued several goals. We have implemented the nuclear-MQM - 
electronic-magnetic-field interaction constant $W_M$ as an expectation value into our four-component 
relativistic correlated all-electron approaches. Section \ref{SEC:THEORY} documents the underlying
theory, details of the implementation, as well as a brief account of the electronic-structure
method for obtaining the required molecular wavefunctions. In section \ref{SEC:APPL} we present
the first detailed theoretical spectroscopic study of the TaN molecule including leading relativistic
effects described in the framework of four-component Dirac theory. Apart from spectroscopic constants
for low-lying electronic states we also discuss computed molecular electric dipole moments and transition
dipole moments. Furthermore, we present and discuss different ${\cal{P,T}}$-odd interaction constants 
relevant to the search for molecular EDMs, as well as the parallel magnetic hyperfine interaction
constant for the {$^{181}$Ta} nucleus with spin $I = 7/2$. The final section \ref{SEC:SUMM} is
dedicated to conclusions drawn from our present findings.


\section{Theory and Methods}
\label{SEC:THEORY}

\subsection{Correlated Relativistic Wavefunction Method}
\label{SUBSEC:CORR_WF}

The four-component molecular Dirac wavefunctions including the contributions of dynamic electron 
correlation have been obtained using a Configuration Interaction approach implemented in the 
\verb+KR-CI+ module \cite{fleig_gasci2,fleig_gasmcscf,knecht_luciparII} of the \verb+Dirac11+ 
\cite{DIRAC11} program package. Throughout the present study we have used the all-electron Dirac-Coulomb 
Hamiltonian operator in Born-Oppenheimer approximation
\begin{equation}
 \label{EQ:HDCMOL}
 \hat{H}^{DC} = \sum\limits_A \sum\limits_i\, \left[ c(\vec{{\bf{\alpha}}} 
 \cdot \vec{p})_i + \beta_i m_0c^2 + V_{iA}{1\!\!1}_4 \right]
                  + \sum\limits_{i,j>i}\, \frac{1}{r_{ij}}{1\!\!1}_4 + \sum\limits_{A,B>A}\, V_{AB},
\end{equation}
where $V_{iA}$ is the potential-energy operator for electron $i$ in the electric field of nucleus
$A$, ${1\!\!1}_4$ is a unit $4 \times 4$ matrix and $V_{AB}$ represents the potential energy due to the internuclear 
classical electrostatic repulsion of the clamped nuclei.

The molecular wavefunctions are defined as a linear expansion into Slater determinants, the latter being 
formed from the basis of molecular time-reversal paired 4-spinors
(Kramers partners, $\left\{\varphi_i,\varphi_{\overline{i}}\right\}$), inter-related as
$\hat{K} \varphi_i = \varphi_{\overline{i}}$ and $\hat{K} \varphi_{\overline{i}} = -\varphi_i$,
where
\begin{equation}
 \label{EQ_K_MANYBODY}
 \hat{K}(n) := e^{-\frac{\imath}{2}\pi \left( \sum\limits_{j=1}^n\, {\boldsymbol{\sigma}}\otimes{1\!\!1}_2(j) \right) \cdot
                        \vec{e}_y}\, \prod\limits_{j=1}^n\, \hat{K}_0(j),
\end{equation}
where the Pauli matrices are defined for particle $j$ in a cartesian basis and $\hat{K}_0$
is the complex conjugation operator. For the one-fermion case, setting $n=1$ and Taylor expanding the exponential in
Eq. (\ref{EQ_K_MANYBODY}) results in the one-body time-reversal operator
\begin{equation}
 \label{EQ_K_ONEBODY}
 \hat{K} := -\imath {\boldsymbol{\Sigma}}_y \hat{K}_0
\end{equation}
with ${\boldsymbol{\Sigma}}_y = \boldsymbol{1}_2 \otimes {\boldsymbol{\sigma}}_y$ a Dirac operator.

The expansion coefficients of the molecular Kramers pairs in terms of the atomic basis sets are obtained 
from an all-electron Dirac-Coulomb Hartree-Fock (DCHF) calculation.

The Slater determinant expansion for the wavefunction then reads
\begin{equation}
 \label{EQ:CI:EXP}
 \left| \psi_k \right> = \sum\limits_{I=1}^{\rm{dim}{\cal{F}}^t(M,N)}\, c_{kI} \left| ({\cal{S}}{\overline{\cal{T}}})_I \right>
\end{equation}
with ${\rm{dim}{\cal{F}}^t(M,N)}$ the dimension of the truncated $N$-particle Fock-space sector over 
$M$ molecular 4-spinors and expansion coefficients $c_{kI}$ varied by diagonalizing the matrix 
representation of $\hat{H}^{DC}$ in the Slater determinant basis. In this step which takes into account 
the correlation part of the electron-electron interaction, the determinants are represented by strings of 
creation operators in second quantization as
\begin{equation}
 \label{EQ:CI:DET}
 \left| ({\cal{S}}{\overline{\cal{T}}}) \right> = {\cal{S}}^{\dagger} {\cal{\overline{T}}}^{\dagger} \rvac
\end{equation}
where ${\cal{S}}^{\dagger} \rvac = \hat{a}_{{\cal{S}}_1}^{\dagger} \hat{a}_{{\cal{S}}_2}^{\dagger} \ldots 
\hat{a}_{{\cal{S}}_j}^{\dagger} \rvac$ and ${\cal{\overline{T}}}^{\dagger} \rvac = \hat{a}_{{\cal{\overline T}}_1}^{\dagger}
\hat{a}_{{\cal{\overline T}}_2}^{\dagger} \ldots \hat{a}_{{\cal{\overline T}}_{N-j}}^{\dagger} \rvac$ 
and $N$ is the number of explicitly treated electrons.
Axial molecular double-group (in the present case the subgroup $C_{64v}^*$ \cite{loras_knecht_fleig_soo}) and
time-reversal symmetry have been exploited.

\subsection{Nuclear magnetic quadrupole moment interaction}

\subsubsection{Basic theory}

The effective Hamiltonian for the interaction of a nuclear magnetic quadrupole moment with an electronic magnetic field 
is \cite{Flambaum_DeMille_Kozlov2014}
\begin{equation}
 \hat{H}_{\text{MQM}} = -\frac{W_M\, M}{2I\left(2I-1\right)}\, \vec{J}_e \hat{T} \vec{n}
 \label{EQ:HNMQM}
\end{equation}
where $\vec{J}_e$ denotes electronic angular momentum and ${\bf{M}}$ 
is the second-rank nuclear magnetic quadrupole moment tensor \cite{Sushkov_Flambaum_Khriplovich1984} with components
\begin{equation}
 M_{mk} = \frac{3M}{2I (2I-1)}\, T_{mk} 
        = \frac{3}{2}\, \frac{M}{I (2I-1)}\, \left[ I_m I_k + I_k I_m - \frac{2}{3}\, I(I+1) \delta_{mk} \right]
 \label{EQ:NMQM}
\end{equation}
with ${\bf{I}}$ the nuclear angular momentum and $M$ the magnitude of the nuclear MQM. For instance, 
if $I_k = I$ along the internuclear vector $\vec{n}$ and $I_{i \ne k} = 0$ then $T_{kk} = \frac{I}{3}\, \left( 4I - 2 \right)$
and $M_{kk} = M$. The ensuing MQM energy shift results to $\Delta E_{MQM} = -\frac{1}{3}\, W_M M \Omega$ with 
$\Omega := \vec{J}_e \cdot \vec{n}$.

The nuclear MQM has two contributions \cite{khriplovich_JETP1976}. The first contribution is due to a possible nucleon EDM, 
proportional to $d_n < 10^{-25}\, e$\, cm.
The second is due to a ${\cal{P,T}}$-odd nuclear potential, proportional to the Fermi constant, order of magnitude
$G_F \approx 10^{-37}$ eV\, cm$^3$. Due to the smallness of the ensuing energy shift and the typical energy differences
$E_{\Omega} - E_{\Omega'}$ between electronic states in diatomic molecules --- expressed in a basis of Hund's case C
functions ---, on the order of eV, higher-order corrections of the type
\begin{equation}
 E^{(2)}_{\text{MQM}}(\Omega_j) = \sum\limits_{k \ne j}\, 
    \frac{\left| \left< \Omega_j | \hat{H}_{\text{MQM}} | \Omega_k \right> \right|^2}{E^{(0)}_{\Omega_j} - E^{(0)}_{\Omega_k}}
\end{equation}
to $E^{(0)}_{\Omega_j}$ are extremely small and may safely be neglected. Considering the case of a doublet
$\left\{\left|\Omega_j\right>, \left|-\Omega_j\right> \right\}$, ($\Omega>0$), the corresponding electronic energies 
$E(\Omega_j)_+$ and $E(\Omega_j)_-$ are near-degenerate for a free molecule, the splitting due to $\Omega$-doubling
being typically on the order of tenths of inverse centimeters for $\Delta$ states \cite{fleig_pthro}, or less.  
In the presence of an external electric field and assuming perfect polarization \cite{ACME_ThO_eEDM_science2014} of
the molecule, $\left\{\left|\Omega_j\right>, \left|-\Omega_j\right> \right\}$ eigenstates are restored, exhibiting a
convenient (but small) Stark splitting. In this two-dimensional subspace, $\left< \Omega_j \left| \hat{H}_{\text{MQM}}
\right| -\Omega_j \right> = \left< -\Omega_j \left| \hat{H}_{\text{MQM}} \right| \Omega_j \right> = 0$ due to
rotational invariance of $\hat{H}_{\text{MQM}}$ for $\vec{J}_e$ the generator of rotation. Furthermore, due to
\begin{displaymath}
 \left< -\Omega_j \left| \hat{H}_{\text{MQM}} \right| -\Omega_j \right> 
       = \left< \Omega_j \left| \hat{P}^{\dagger} \hat{H}_{\text{MQM}} \hat{P} \right| \Omega_j \right>
       = - \left< \Omega_j \left| \hat{H}_{\text{MQM}} \right| \Omega_j \right>
\end{displaymath}
and
\begin{displaymath}
 \left< -\Omega_j \left| \hat{H}_{\text{MQM}} \right| -\Omega_j \right> 
       = \left< \Omega_j \left| \hat{K}^{\dagger} \hat{H}_{\text{MQM}} \hat{K} \right| \Omega_j \right>^*
       = - \left< \Omega_j \left| \hat{H}_{\text{MQM}} \right| \Omega_j \right>^*
\end{displaymath}
with $\hat{P}$ the operator of spatial inversion, only one matrix element,
$\left< \Omega_j \left| \hat{H}_{\text{MQM}} \right| \Omega_j \right>$, needs to be calculated for the subspace
\cite{kozlov_PVdiat1995}.

\subsubsection{The nuclear-MQM electronic-magnetic-field interaction constant}

Be $M_{nk}$ a component of the magnetic quadrupole moment due to a nuclear charged current. Then the $i$-th
component of the quadrupole term of the associated classical vector potential is written as
\cite{Sushkov_Flambaum_Khriplovich1984}
\begin{eqnarray}
 A_Q(\vec{r})_{\imath} &=& -\frac{1}{6}\, \sum\limits_{k,l,n}\, \varepsilon_{\imath ln}\, M_{nk}\, \frac{\partial}{\partial r_l}
                                                \frac{\partial}{\partial r_k}\, \frac{1}{r}
         = \frac{1}{6}\, \sum\limits_{k,l,n}\, \varepsilon_{\imath ln}\, M_{nk}\,
                          \left( \frac{\delta_{kl}}{r^3} - \frac{3 r_k r_l}{r^5} \right)
\end{eqnarray}
for positions $\vec{r}$ outside of the nuclear current distribution. Since $M$ is a symmetric tensor, the nuclear vector 
potential thus becomes
\begin{equation}
 \vec{A}_Q(\vec{r}) = -\sum\limits_{k,n}\, M_{nk}\, \frac{1}{2r^5}\, 
              \sum\limits_{\imath,l}\, \varepsilon_{\imath ln}\, r_l r_k \vec{e}_{\imath}
\end{equation}
The potential energy due to an electron moving with velocity $\vec{v}$ relative to the origin of this vector potential,
$V_{Qe} = -\frac{e}{c}\, \vec{v} \cdot \vec{A}_Q(\vec{r})$ can now be quantized for a Dirac particle via
$\vec{v} \rightarrow c\vec{\alpha}$, yielding the corresponding Hamiltonian operator, in atomic units
\begin{equation}
 \label{EQ:HQ_HAMIL}
 \hat{H}_{Qe} = \sum\limits_{k,n}\, M_{nk}\, \frac{1}{2r^5}\, 
                \sum\limits_{\imath,j,l}\, \varepsilon_{\imath ln}\, r_l r_k\, \vec{e}_{\imath} \cdot \vec{e}_j\, \alpha_j
              = \sum\limits_{j,k,l,n}\, \frac{1}{2r^5}\, \varepsilon_{jln}\, \alpha_j\, r_l r_k\, M_{kn}
\end{equation}
Now, introducing a contracted tensor $(\vec{rM}) = \sum\limits_{\imath}\, \vec{e}_{\imath}\, \left(rM\right)_i$,
Eq. (\ref{EQ:HQ_HAMIL}) can be rewritten as
\begin{equation}
 \hat{H}_{Qe} = \sum\limits_{j,l,n}\, \frac{1}{2r^5}\, \varepsilon_{jln}\, \alpha_j\, r_l\, \vec{e}_n \cdot \left(\vec{rM}\right)
              = \frac{\vec{\alpha} \times \vec{r}}{2r^5}\, \cdot \left(\vec{rM}\right).
 \label{EQ:HQ_HAMIL2}
\end{equation}
%
This Hamiltonian can be introduced as a perturbation operator for a single Dirac particle interacting with a nuclear
MQM. With regard to Eqs. (\ref{EQ:HNMQM}) and (\ref{EQ:NMQM}), and Eq. (\ref{EQ:HQ_HAMIL2}) the nuclear-MQM 
electronic-magnetic-field interaction constant is defined for a many-electron molecular system with axial symmetry as
\begin{equation}
 W_M := \frac{3}{2\Omega}\, \left< \Psi_{\Omega} \left| \sum\limits_{j=1}^n\,
                \left( \frac{\boldsymbol{\alpha}_j \times {\bf{r}}_{jA}}{r_{jA}^5} \right)_k\,
                \left( r_{jA} \right)_k \right| \Psi_{\Omega} \right>
 \label{EQ:MQM_INTCONST}
\end{equation}
for nucleus $A$, where $j$ is an electron index, $k$ a cartesian component, $\boldsymbol{\alpha}$
is a Dirac matrix, and $\Omega = \left< {\bf{J}}_{\text{e}} \cdot {\bf{n}} \right>$ is the projection
of the total electronic angular momentum ${\bf{J}}_{\text{e}}$ onto the molecular axis ${\bf{n}}$.
In the present case, the nuclei of Ta and N are placed along the $z$ axis, and so ${\bf{n}} = {\bf{e}}_z$.

With the wavefunctions defined in subsection \ref{SUBSEC:CORR_WF} the evaluation of the MQM interaction
constant can be written as
\begin{equation}
 W_M(\Psi_k) = \frac{3}{2\Omega}\, \sum\limits_{I,J=1}^{\rm{dim}{\cal{F}}^t(M,N)}\, c^*_{kI}c_{kJ}\, 
                      \left< ({\cal{S}}{\overline{\cal{T}}})_I \right| 
    \sum\limits_{j=1}^n\,
                \left( \frac{\boldsymbol{\alpha}_j \times {\bf{r}}_{jA}}{r_{jA}^5} \right)_l\,
                \left( r_{jA} \right)_l \left| ({\cal{S}}{\overline{\cal{T}}})_J \right>
 \label{EQ:CIEXPECVAL}
\end{equation}
with optimized CI coefficients $\left\{c_{kM}\right\}$ for the $k$th $N$-electron state $\psi_k$ in
irreducible representation $\Omega$.

The perturbative operator in Eq. (\ref{EQ:CIEXPECVAL}) is a manifestly ${\cal{P,T}}$-odd one-electron
operator that in a basis of time-reversal paired molecular spinors can be written in second quantization as
\begin{equation}
 \hat{H}^M = \sum\limits_{p,q=1}^{P_u}\, h^M_{pq}\, a_p^{\dagger} a_q
            +\sum\limits_{p=1}^{P_u} \sum\limits_{q = P_u+1}^P\, h^M_{p\overline{q}}\, a_p^{\dagger} a_{\overline{q}}
            +\sum\limits_{p=P_u+1}^P \sum\limits_{q = 1}^{P_u}\, h^M_{\overline{p}q}\, a_{\overline{p}}^{\dagger} a_q
            +\sum\limits_{p,q=P_u+1}^P h^M_{\overline{p}\overline{q}}\, a_{\overline{p}}^{\dagger} a_{\overline{q}}
 \label{EQ:HMQMSECQUANT}
\end{equation}
where $P_{u(b)}$ is the number of Kramers unbarred (barred) spinors in the molecular basis set and $P = P_u + P_b$.
We write Eq. (\ref{EQ:CIEXPECVAL}) using the first term in Eq. (\ref{EQ:HMQMSECQUANT}) to illustrate the computation
of the expectation value:
{\small{
\begin{equation}
 W_M(\Psi_k)_1 = \frac{3}{2\Omega}\, \sum\limits_{I,J=1}^{\rm{dim}{\cal{F}}^t(P,N)}\, c^*_{kI}c_{kJ}\,
                 \sum\limits_{m,n=1}^{P_u}\, h^M_{mn}\, 
                 \lvac \prod\limits_{p=1}^{N_p \in {\cal{S}}_I} \prod\limits_{\overline{p} = N_p+1}^{N_p\in{\cal{S}}_I 
                       + N_{\overline{p}}\in{\cal{\overline{T}}}_I}\, 
                       a_{\overline{p}} a_p\, a_m^{\dagger} a_n
                       \prod\limits_{q=1}^{N_p\in{\cal{S}}_J} \prod\limits_{\overline{q} = N_p+1}^{N_p\in{\cal{S}}_J 
                       + N_{\overline{p}}\in{\cal{\overline{T}}}_J}\, 
                       a^{\dagger}_q a^{\dagger}_{\overline{q}}\, \rvac
 \label{EQ:WMSECQUANT1}
\end{equation}
}}
with $N_{p({\overline{p}})}$ the number of electrons occupying the string ${\cal{S}}_K ({\cal{\overline{T}}}_K)$ and 
$N = N_p + N_{\overline{p}}$ in a given string combination (see Eq. (\ref{EQ:CI:DET})) forming a Slater determinant. 
The interaction constant thus results from the contraction of CI densities, molecular integrals and coupling coefficients.

\subsubsection{Implementation}

The one-electron operator in Eq. (\ref{EQ:CIEXPECVAL}) can be written more explicitly for the
component along the molecular axis (\ie\ $l = z$) using
\begin{equation}
 \left( \frac{\boldsymbol{\alpha} \times {\bf{r}}}{r^5} \right)_z\, r_z 
    = \alpha_1\, \frac{r_2 r_3}{r^5} - \alpha_2\, \frac{r_1 r_3}{r^5}
 \label{EQ:MQM_OPEREXPL}
\end{equation} 
with $r_i$ the $i$th cartesian component of the electronic position vector $\bf{r}$.
The components of the electric-field gradient, a second-rank tensor, are
\begin{equation}
 \frac{1}{q}\, \frac{\partial}{\partial r_i}\, E_j({\bf{r}}) = -3 \frac{r_i r_j}{r^5}
 \label{EQ:MQM_EFG}
\end{equation}
for a point charge $q$ and $i \ne j$.
We calculate atomic integrals over the two terms on the right-hand side of Eq. 
(\ref{EQ:MQM_OPEREXPL}) as integrals over corresponding components of electric field gradients, as
\begin{equation}
 \iiint_{\cal{V}}\, \frac{r_i r_j}{r^5} d^3r 
     = -\frac{1}{3}\, \iiint_{\cal{V}}\, \frac{\partial}{\partial r_i}\, \frac{r_j}{r^3}\, d^3r.
 \label{EQ:EFG_INT}
\end{equation}
The non-vanishing integrals over the full four-spinors are then those of the generic
type
\begin{equation}
 \left< \Psi^L \left| \sigma_k\, \frac{r_i r_j}{r^5} \right| \Psi^S \right>
 \label{EQ:EFG_INT_ALPHA}
\end{equation}
with $\sigma_k$ the $k$th spin-Pauli matrix and the small-component wavefunction $\Psi^S$ in the
ket-vector of the same spatial parity as the large-component wavefunction $\Psi^L$ in the bra-vector, 
since the $\alpha$ matrices couple large and small components of the four-spinors. This can only be
the case if $\Psi^L$ and $\Psi^S$ come from different atomic contributions to a molecular spinor,
reflecting the fact that the molecular spinors are not parity eigenfunctions.

After the computation of these integrals in the atomic spinor basis, an integral transformation is
performed into the molecular spinor basis using the expansion coefficients from a preceding 
DCHF calculation, yielding one-electron integrals of the type $h^M_{pq}$ in Eq. 
(\ref{EQ:WMSECQUANT1}).

\subsubsection{Other interaction constants}

The electron EDM interaction constant is evaluated in accord with stratagem II of Lindroth et al.
\cite{lindroth_EDMtheory1989} as an effective one-electron operator via the squared electronic 
momentum operator,
\begin{equation}
 W_d := \frac{2\imath c}{\Omega\, e\hbar} \left<
          \sum\limits_{j=1}^n\, \gamma^0_j \gamma^5_j\, \vec{p}_j\,^2 \right>_{{\psi}_{\Omega}}
\end{equation}
with $n$ the number of electrons, as described in greater detail in reference \cite{fleig_nayak_eEDM2013}. 
It is related to the EDM effective electric field as $E_{\text{eff}} = \Omega\, W_d$. 

The eN-SPS interaction constant is defined and implemented \cite{ThF+_NJP_2015} as
\begin{equation}
 W_{\cal{P},\cal{T}} := \frac{\imath}{\Omega}\,
          \frac{G_F}{\sqrt{2}}\, Z\, \left<\sum\limits^n_{j=1}\, 
                     {\gamma^0_j\gamma^5_j\, \rho_N(\vec{r}_j)}
                    \right>_{{\psi}_{\Omega}}
\end{equation}
where $G_F$ is the Fermi constant, $Z$ is the proton number and $\rho_N(\vec{r}_j)$ is the nuclear charge
density at position $\vec{r}_j$, normalized to unity.

In addition to the above ${\cal{P,T}}$-odd interaction constants we investigate in the present work
magnetic hyperfine interaction which serves as an evaluation quantity for the quality of electronic
wavefunctions, since spin density close to nuclei is probed. The parallel hyperfine constant is
defined as follows:
\begin{equation}
  A_{||} = \frac{\mu_{A}}{I \Omega}\, 
             \left< \sum\limits_{i=1}^n\, \left( \frac{\vec{\alpha_i} \times \vec{r}_{iA}}{r_{iA}^3}
                               \right)_z \right>_{{\psi}_{\Omega}}
\end{equation}
where $A$ designates a nucleus, $\vec{\alpha}$ is a vector of Dirac matrices, and $\vec{r}_{iA}$
is the position vector relative to nucleus $A$. Further details can be found in reference \cite{Fleig2014}.

\section{Application to TaN}
\label{SEC:APPL}

\subsection{Technical details}

\subsubsection{General}
A local version of the \verb+DIRAC11+ program package \cite{DIRAC11} has been used for all presented
calculations. This version has been extended to allow for the calculation of expectation values over
the various property operators reported in this work (see references 
\cite{fleig_nayak_eEDM2013,Fleig2014,ThF+_NJP_2015}), in the present case the nuclear magnetic 
quadrupole moment interaction. Correlated wavefunctions were obtained with the \verb+KR CI+ module
\cite{knecht_luciparII}. In all calculations the speed of light was set to 137.0359998 a.u.

\subsubsection{Atomic basis sets}
Fully uncontracted all-electron atomic Gaussian basis sets of triple-$\zeta$ quality were used for 
the description of electronic shells. For tantalum we used Dyall's basis set 
\cite{dyall_basis_2004,dyall_gomes_basis_2010} and for nitrogen the Dunning cc-pVTZ-DK set 
\cite{Dunning_jcp_1989}. For tantalum valence- and core-correlating exponents were added to the 
basic triple-$\zeta$ set, amounting to \{$30s,24p,15d,11f,3g,1h$\} uncontracted functions.
In electron-correlated calculations the virtual spinor space has been consistently truncated at 
$30$ a.u.

\subsubsection{Molecular spinors}
Kramers-paired four-spinors have been obtained from Dirac-Coulomb Hartree-Fock calculations including
4-index integrals (unapproximated) over Small spinor components. The chosen model used a Fock operator
where occupation numbers were averaged by evenly distributing $2$ electrons over the valence
spinors with the leading character Ta($s_{\sigma}$) and a pair of Ta($d_{\delta}$). Such a basis set of 
molecular spinors is expected to yield a balanced description of ground $\Sigma$ and first excited
$\Delta$ states of the TaN molecule while being very similar to excited-state specific $\Delta$
spinors. For comparison, we have also tested the latter by restricting one electron to Ta($s_{\sigma}$)
and one electron to Ta($d_{\delta}$) and call them $\Delta$-spinors.

\subsubsection{Configuration spaces}
\label{SUBSUBSEC:CONF_SPACES}

\begin{figure}[h]
 \caption{\label{FIG:TAN_GAS}
          Generalized Active Space models for TaN CI wavefunctions. The parameters $m, n, p, q$
          are defined in subsection \ref{SUBSUBSEC:CONF_SPACES} and determine the occupation 
          constraints of the
          subspaces of Kramers-paired spinors. The molecular spinors are denoted according to
          their principal atomic character. The space with $110$ virtual Kramers pairs
          is comprised by all canonical DCHF spinors below an energy of $30$ E$_H$.
         }

\vspace{0.5cm}

 \begin{center}
  \includegraphics[width=10.0cm,angle=0]{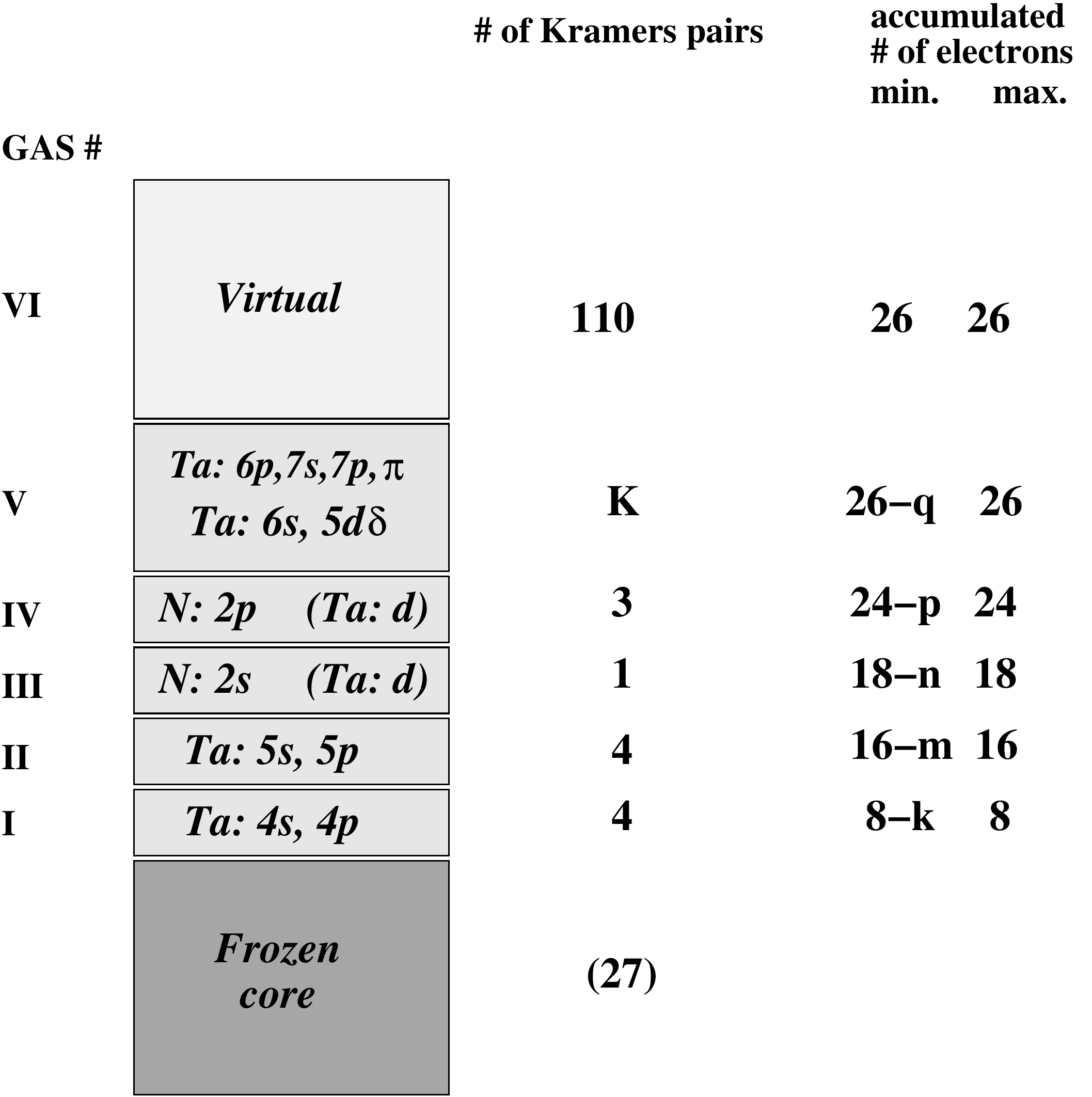}
 \end{center}
\end{figure}

We exploit the Generalized Active Space (GAS) concept for defining CI wavefunctions of varying
quality. Figure \ref{FIG:TAN_GAS} shows the partitioning of the space of Kramers-paired
spinors into seven subspaces, six of which are active for replacements of either particle or
hole type. Based on this

partitioning and five parameters ($k, m, n, p, q$) which define the accumulated occupation
constraints of the subspaces, we choose a number of different CI wavefunction models for our 
calculations:
\begin{center}
 \begin{tabular}{l|c|r}
  parameter values       & correlation model acronym  & space dimension ($\Omega=1$)  \\ \hline
  $k=0, m=0, n=2, p=2, q=2$   & MR$_K$-CISD(10)            &    $24,467,439$  \\
  $k=0, m=0, n=2, p=2, q=3$   & MR$_K$-CISDT(10)           &   $156,703,118$  \\
  $k=0, m=0, n=2, p=3, q=2$   & MR$^{+T}_K$-CISD(10)       &   $345,490,884$  \\
  $k=0, m=1, n=2, p=2, q=2$   & MR$^{'}_K$-CISD(18)        &    $79,033,385$  \\
  $k=0, m=2, n=2, p=2, q=2$   & MR$_K$-CISD(18)            &   $102,823,317$  \\
  $k=0, m=2, n=2, p=2, q=3$   & MR$_K$-CISDT(18)           &   $691,737,264$  \\
  $k=0, m=1, n=2, p=3, q=2$   & MR$^{'+T}_K$-CISD(18)      & $1,616,196,776$  \\
  $k=0, m=2, n=2, p=3, q=2$   & MR$^{+T}_K$-CISD(18)       & $2,581,590,063$  \\
  $k=1, m=1, n=2, p=2, q=2$   & MR$^{'}_K$-CISD(26)        &   $133,599,331$ 
 \end{tabular}
\end{center}
The parameter $K$ (see Fig. \ref{FIG:TAN_GAS}) has been introduced to define the active valence
spinor spaces, and the dimension of the CI space is given for $K=12$. $K=3$ ($1,492,556$ 
determinants for $\Omega=1$ in the model MR$_3$-CISD(10)) includes only the Ta $(6s,5d_{\delta})$ 
spinors in the fifth active space and thus comprises a minimal model for a balanced description 
of the ground $\Omega=0$ state and the excited $\Omega\in\{1,2,3\}$ states which derive from 
${^3\Delta}$ and ${^1\Delta}$ in the $\Lambda$-$S$ coupling picture. 
The different wavefunction models are in addition defined by the number of correlated electrons in
total (in parenthesis) and the included maximum excitation ranks, where ``SDT'' stands for Single, 
Double, and Triple excitations, as an example. 


\subsubsection{Hyperfine interaction}
For the determination of the nuclear magnetic hyperfine coupling constant we use the tantalum isotope
$\rm {^{181}{T}a}$ for which the nuclear magnetic moment has been determined to be
$\mu = 2.361 \mu_N$ \cite{Ta181_magmom_1973}. Its nuclear spin quantum number is $I = 7/2$.


\subsection{Results and discussion}

\subsubsection{Spectroscopic properties}
\label{SUBSUBSEC:SPECPROP}

Table \ref{TAB:TAN:ACTIVESPINORS} displays character and some properties of the active set of spinors.

\begin{table}[h]

\caption{\label{TAB:TAN:ACTIVESPINORS}
         Characterization of active-space Kramers pairs in terms of orbital angular momentum projection,
         spinor energy, and principal atomic shell character; the Kramers pairs numbered 29-35 represent
         the Ta(4f) shell.
        }

\begin{center}
\begin{tabular}{c|ccrrl}
 No. & \rule{0.3cm}{0.0cm} $|m_j|$ & \rule{0.7cm}{0.0cm} $\left|\left< \hat{\ell}_z \right>_{\varphi_i}\right|$
                    & \rule{0.5cm}{0.0cm} $\varepsilon_{\varphi_i}$ [$E_H$]
                    & \rule{0.5cm}{0.0cm} atomic population, Atom($\ell_{\lambda}$) \\ \hline
 25 &  $1/2$  & $0.000$   & $-3.184$ & \rule{0.5cm}{0.0cm} $99$\% Ta($s$)       \\
 26 &  $1/2$  & $0.623$   & $-2.016$ & \rule{0.5cm}{0.0cm} $36$\% Ta($p_{\sigma}$), $62$\% Ta($p_{\pi}$)        \\
 27 &  $1/2$  & $0.374$   & $-1.697$ & \rule{0.5cm}{0.0cm} $57$\% Ta($p_{\sigma}$), $37$\% Ta($p_{\pi}$)        \\
 28 &  $3/2$  & $1.000$   & $-1.668$ & \rule{0.1cm}{0.0cm} $99$\% Ta($p_{\pi}$)        \\
 $\vdots$ & $\vdots$ & $\vdots$ & $\vdots$ & $\vdots$ \\
 36 &  $1/2$  & $0.003$   & $-0.879$ & \rule{0.1cm}{0.0cm} $81$\% N($s$), $9$\% Ta($d_{\sigma}$)  \rule{0.0cm}{0.8cm}       \\
 37 &  $1/2$  & $0.287$   & $-0.362$ & \rule{0.5cm}{0.0cm} $46$\% N($p_{\sigma}$), $17$\% N($p_{\pi}$), $14$\% Ta($d_{\sigma}$), $10$\% Ta($d_{\pi}$) \\
 38 &  $1/2$  & $0.713$   & $-0.353$ & \rule{0.5cm}{0.0cm} $43$\% N($p_{\pi}$), $24$\% Ta($d_{\pi}$), $18$\% N($p_{\sigma}$) \\
 39 &  $3/2$  & $1.001$   & $-0.353$ & \rule{0.5cm}{0.0cm} $61$\% N($p_{\pi}$), $35$\% Ta($d_{\pi}$) \\ \hline
 40 &  $1/2$  & $0.002$   & $-0.095$ & \rule{0.5cm}{0.0cm} $79$\% Ta($s$), $10$\% Ta($d_{\sigma}$), $8$\% Ta($p_{\sigma}$)        \\
 41 &  $3/2$  & $1.996$   & $-0.017$ & \rule{0.5cm}{0.0cm} $100$\% Ta($d_{\delta}$)        \\
 42 &  $5/2$  & $2.000$   & $-0.007$ & \rule{0.5cm}{0.0cm} $100$\% Ta($d_{\delta}$)        \\ \hline
 43 &  $1/2$  & $0.990$   & $ 0.022$ & \rule{0.5cm}{0.0cm} $88$\% Ta($p_{\pi}$)        \\
 44 &  $3/2$  & $1.001$   & $ 0.025$ & \rule{0.5cm}{0.0cm} $90$\% Ta($p_{\pi}$)        \\
 45 &  $1/2$  & $0.008$   & $ 0.041$ & \rule{0.5cm}{0.0cm} $60$\% Ta($p_{\sigma}$), $14$\% Ta($s$), $13$\% Ta($d_{\sigma}$)       \\
 46 &  $1/2$  & $0.023$   & $ 0.094$ & \rule{0.5cm}{0.0cm} $86$\% Ta($s$), $13$\% Ta($d_{\sigma}$)       \\
 47 &  $1/2$  & $0.636$   & $ 0.103$ & \rule{0.5cm}{0.0cm} $51$\% Ta($p_{\pi}$), $25$\% Ta($p_{\sigma}$), $10$\% Ta($d_{\pi}$)  \\
 48 &  $3/2$  & $1.001$   & $ 0.113$ & \rule{0.5cm}{0.0cm} $77$\% Ta($p_{\pi}$)      \\
 49 &  $1/2$  & $0.342$   & $ 0.115$ & \rule{0.5cm}{0.0cm} $36$\% Ta($p_{\sigma}$), $26$\% Ta($p_{\pi}$), $16$\% Ta($d_{\sigma}$)  \\
 50 &  $3/2$  & $1.030$   & $ 0.240$ & \rule{0.5cm}{0.0cm} $47$\% Ta($d_{\pi}$), $34$\% Ta($p_{\pi}$), $13$\% N($p_{\pi}$)      \\
 51 &  $1/2$  & $1.000$   & $ 0.240$ & \rule{0.5cm}{0.0cm} $47$\% Ta($d_{\pi}$), $34$\% Ta($p_{\pi}$), $16$\% N($p_{\pi}$)   
\end{tabular}

\end{center}
\end{table}

The orbital angular momentum projection quantum number $\lambda$ has been assigned based on a
Mulliken population analysis, and the respective mixings of different $\lambda$ values in a given
Kramers pair are correctly reflected by the $\hat{\ell}_z$ expectation values.

Of particular interest is the character of the $\sigma_{6s_{\text{Ta}}}$ spinor, which displays
79\% $6s_{\text{Ta}}$, 8\% $6p_{z\text{Ta}}$, and 10\% $5d_{x^2,y^2,z^2\text{Ta}}$ leading atomic
contributions. Since reference \cite{Flambaum_DeMille_Kozlov2014} draws a comparison with the YbF
molecule for estimating ${\cal{P,T}}$-odd matrix elements, we have performed similar DCHF
calculations on YbF. Here, the $\sigma_{6s_{\text{Yb}}}$ spinor has 84\% $6s_{\text{Yb}}$, 
13\% $6p_{z\text{Yb}}$, and a few \% $5d_{x^2,y^2,z^2\text{Ta}}$ contribution, confirming the
assumption made by Flambaum et al. \cite{Flambaum_DeMille_Kozlov2014} that the $p$-wave contribution
is significantly smaller in TaN. It is noteworthy that our finding is not in accord with the 
theoretical results of Bernath et al. \cite{TaN_Bernath_2002} on this point, where no $p$-wave
contribution has been found in the case of TaN. However, our present DCHF calculations account for
spin-orbit interaction in contrast to the scalar relativistic calculations in reference
\cite{TaN_Bernath_2002}. 

Earlier internally contracted scalar-relativistic MRCI calculations \cite{TaN_Bernath_2002} on TaN 
electronic states have shown that the wavefunction of the excited $^3\Delta$ state which is of particular 
interest in the present study contains non-negligible contributions from two types of Triple excitations 
which we denote using the Kramers pair numbering indices from Table \ref{TAB:TAN:ACTIVESPINORS}:
\begin{eqnarray*}
  (38,39)^2 (40)^1        &\longrightarrow& (41,42)^1 (43,44)^2 \\
  (37)^1 (38,39)^1 (40)^1 &\longrightarrow& (41,42)^1 (43,44)^1 (45)^1
\end{eqnarray*}
Here, the superscript denotes an occupation number.
Our CI expansions of the type MR$^{+T}_{12}$-CISD, therefore, include an important subset of Quintuple 
excitations with respect to the $^1\Sigma$ ground state of TaN which constitute Double excitations with 
respect to the above triply-excited configurations. These Double excitations are written using the GAS 
notation from Fig. \ref{FIG:TAN_GAS} as
\begin{eqnarray*}
 \text{V}^2 &\longrightarrow& \text{VI}^2 \\
 \text{IV}^1 \text{V}^1 &\longrightarrow& \text{VI}^2 \\
 \text{III}^1 \text{V}^1 &\longrightarrow& \text{VI}^2.
\end{eqnarray*}
Table \ref{TAB:TAN:VERTEXCIT} shows vertical excitation energies for five low-lying electronic states.

\begin{table}[h]

\caption{\label{TAB:TAN:VERTEXCIT}
         Calculated vertical excitation energies T$_v$ for $\Omega \in \{0,1,2,3\}$ at $R = 3.1806$ a$_0$
        }

\begin{center}
\begin{tabular}{c|rrrrr}
                  & \multicolumn{5}{c}{Correlation model}                                                             \\
 State ($\Omega$) & CAS2      & CAS-SD2 & MR$_3$-CISD($10$) & MR$_{12}$-CISD($10$) & MR$^{+T}_{12}$-CISD($10$)  \\ \hline
    $2$           & $13409$   & $9164$  &    $11321$        &    $11749$           &     $11693$                \\
    $3$           & $2896$    & $3754$  &     $3436$        &     $5259$           &      $5434$                \\
    $2$           & $1214$    & $1977$  &     $1801$        &     $3597$           &      $3772$                \\
    $1$           & $   0$    & $ 917$  &     $ 682$        &     $2502$           &      $2673$                \\
    $0$           & $3147$    & $   0$  &     $   0$        &     $   0$           &      $   0$                 
\end{tabular}
\end{center}
\end{table}

CAS2 denotes a 2-electron Full CI calculation in the $\sigma_{6s_{\text{Ta}}}$, $\delta_{5d_{\text{Ta}}}$
subspace, and CAS2-SD2 a 2-electron Full CI calculation including all virtual spinors up to an energy
of 30 a.u. It becomes clear that an active spinor space including
only the three above-mentioned spinor pairs is insufficient in describing correctly the relative energies
of the low-lying states. We, therefore, base all following studies on models of the type MR$_{12}$.

In Table \ref{TAB:TAN:SPECCON} we compare calculated spectroscopic constants from different electronic
structure models with experiment.

\begin{table}[h]

\caption{\label{TAB:TAN:SPECCON}
         Calculated spectroscopic constants ($R_e$ the equilibrium internuclear distance, $\omega_e$ the harmonic vibrational
         frequency, $B_e$ the rotational constant, and $T_e$ the equilibrium excitation energy) for $\Omega \in \{0,1,2,3\}$ 
         and comparison with experimental values where available
        }

\begin{center}
\begin{tabular}{ll|cccc}
 State            & Model                    & $R_e$ [a.u.] & $\omega_e$ [\cm]  & $B_e$ [\cm]   &  $T_e$ [\cm]   \\ \hline
                  & av. DCHF                 &  $3.115$     & $1163$                   &  $0.477$    &       $0$      \\ \hline
 ${^1\Sigma}^+_0$ & MR$_{12}$-CISD($10$)     &  $3.160$     & $1161$                   &  $0.464$    &       $0$      \\
                  & MR$^{+T}_{12}$-CISD($10$)&  $3.205$     & $1122$                   &  $0.451$    &       $0$      \\
                  & MR$_{12}$-CISD($18$)     &  $3.136$     & $1184$                   &  $0.471$    &       $0$      \\
                  & MR$_{12}$-CISD($18$)+T   &  $3.181$     & $1134$                   &  $0.458$    &       $0$      \\
                  & Exp.                     &  $3.181$\footnotemark[2] & $1070$\footnotemark[1]     &   &  $0.0$      \\ \hline
 ${^3\Delta}_1$   & MR$_{12}$-CISD($10$)     &  $3.170$     & $1116$                   &  $0.461$    &    $2526$      \\
                  & MR$^{+T}_{12}$-CISD($10$)&  $3.222$     & $1088$                   &  $0.446$    &    $2598$      \\
                  & MR$_{12}$-CISD($18$)     &  $3.148$     & $1136$                   &  $0.468$    &    $2879$      \\
                  & MR$_{12}$-CISD($18$)+T   &  $3.196$     & $1095$                   &  $0.454$    &    $2967$      \\
                  & Exp.                     &  $3.196$\footnotemark[2] &       &      &  $2827.2917$\footnotemark[2] \\ \hline
 ${^3\Delta}_2$   & MR$_{12}$-CISD($10$)     &  $3.169$     & $1117$                   &  $0.461$    &    $3618$      \\
 ${^3\Delta}_3$   & MR$_{12}$-CISD($10$)     &  $3.168$     &      $1119$              &  $0.462$    &    $5276$      \\
 ${^1\Delta}_2$   & MR$_{12}$-CISD($10$)     &  $3.153$     &      $1123$              &  $0.466$    &   $11729$      \\ \hline
\end{tabular}

\footnotetext[1]{Reference \cite{andrews_JPCA1998}}
\footnotetext[2]{T$_0$ from Reference \cite{TaN_Bernath_2002}}

\end{center}
\end{table}

\begin{figure}[h]
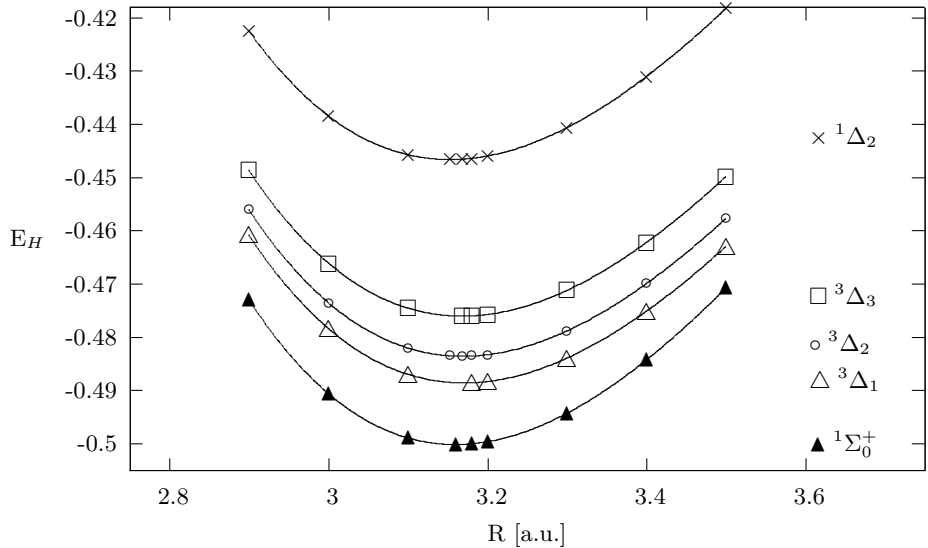

 \caption{\label{FIG:TaN_LOWSTATES}
          Potential energy curves for five lowest-lying electronic states of TaN, using the CI model MR$_{12}$-CISD(10), 
          cutoff $30$ a.u.  The energy offset is $-15671$ E$_H$.
         }

\vspace{0.5cm}

 \begin{center}
\setlength{\unitlength}{0.240900pt}
\ifx\plotpoint\undefined\newsavebox{\plotpoint}\fi
\sbox{\plotpoint}{\rule[-0.200pt]{0.400pt}{0.400pt}}%

 \end{center}
\end{figure}

Potential-energy curves for the four lowest-lying electronic states
are displayed in Figure \ref{FIG:TaN_LOWSTATES}.

The additional selected Triple excitations stretch the lower-lying
potential energy curves significantly, presumably owing to partial depopulation of bonding spinors.
This stretching is, not surprisingly, accompanied by a decrease in harmonic vibrational frequency. The
opposite is observed when including determinants with one and two holes in the $5s_{\text{Ta}}$ and
$5p_{\text{Ta}}$ spinors in the wavefunction expansion. This partial depopulation of compact tantalum
outer-core shells results in an increased effective nuclear electric potential ``seen'' by the bonding 
valence electrons and is therefore the likely reason for the bond contraction. Both effects are
important, and we therefore deduce constants for the lowest two states from the 18-electron model with 
energies of the individual Born-Oppenheimer points corrected by the energy due to the Triples correction 
in the 10-electron model. This corrected model is called MR$_{12}$-CISD($18$)+T and yields very accurate 
equilibrium bond lengths. There remains a deviation from experiment of more than 60 \cm\ for the 
ground-state harmonic vibrational frequency, however.

Since no experimental data are available for the rotation constant, we only mention here that the rotation
constant is largely insensitive to different electronic-structure models, the value varying on the
order of a few percent.

The equilibrium excitation energy T$_e$ for the first excited ($\Omega=1$) state is already in quite
good agreement with the experimental value for the MR$^{+T}_{12}$-CISD($10$) model. Including outer-core
correlations, however, lead to a substantial further improvement, with the MR$_{12}$-CISD($18$)+T model
overshooting by only about 170 \cm.

Since active-space Triple excitations make an important contribution to the static molecular dipole moment
of the $\Omega=1$ state (see Table \ref{TAB:TAN:PROPER}), we present in Table \ref{dipolemoments} 
molecular dipole moments and transition dipole moments computed with the model MR$^{+T}_{12}$-CISD($10$).

\begin{table}[h]
 \caption{Molecular static electric dipole moments
        $\left<{^M\Lambda}_{\Omega} | {\hat{D}}_z | {^M\Lambda}_{\Omega} \right>$,
          transition dipole moments
        $\left|\left|\left<{^M\Lambda}_{\Omega}' | {\hat{\vec{D}}} | {^M\Lambda}_{\Omega} \right>\right|\right|$,
          with ${\hat{\vec{D}}}$ the electric dipole moment operator (both in Debye units),
          using the model MR$^{+T}_{12}$-CISD($10$). The origin is at the center of mass, and the internuclear 
          distance is $R=3.1806$ [$a_0$] (N nucleus placed at $z\vec{e}_z$ with $z>0$).
         }
\label{dipolemoments}

\vspace*{0.5cm}
\hspace*{-2.8cm}
\begin{tabular}{c|lllll}
 ${{^M\Lambda}_{\Omega}}$ State
                                  & \rule{0.1cm}{0.0cm} ${^1}\Sigma^+_0$ \rule{0.5cm}{0.0cm}
                                  & ${^3}\Delta_1$   \rule{0.5cm}{0.0cm}
                                  & ${^3}\Delta_2$   \rule{0.5cm}{0.0cm}
                                  & ${^3}\Delta_3$   \rule{0.5cm}{0.0cm}
                                  & ${^1}\Delta_2$   \rule{0.0cm}{0.0cm}     \\ \hline
 ${^1}\Sigma^+_0$  &\rule{0.1cm}{0.0cm} $-3.515$    &     &     &     &      \\
 ${^3}\Delta_1$    &\rule{0.1cm}{0.0cm} $0.028$    & $-4.809$  &     &     &      \\
 ${^3}\Delta_2$    &\rule{0.1cm}{0.0cm} $0.000$    & $0.085$  & $-4.775$  &     &      \\
 ${^3}\Delta_3$    &\rule{0.1cm}{0.0cm} $0.000$    & $0.000$  & $0.087$  & $-4.776$  &      \\
 ${^1}\Delta_2$    &\rule{0.1cm}{0.0cm} $0.000$    & $0.139$  & $0.114$  & $0.164$  & $-4.000$
\end{tabular}
\end{table}

In the state that will possibly be used for a future EDM measurement, $\Omega=1$, the molecule has a
very large electric dipole moment of about 4.8 Debye. This is for one thing an asset regarding the 
polarisation of the molecule along an external electric laboratory field. For the other, a large molecular
dipole moment also implies a large EDM effective electric field which will be discussed in the following
subsection. Transition dipole moments have the expected qualitative agreement with selection rules for
electric dipole transitions. The value for $\left|\left|\left< {^3}\Delta_1 | {\hat{\vec{D}}} | 
{^1}\Sigma^+_0 \right>\right|\right| \approx 0.03$ Debye is of particular interest. Since 
$\Delta\Lambda = \pm 2$ for this transition, it should be electric-dipole forbidden. The non-zero value 
for the transition dipole moment indicates that $\Lambda = 2$ is not an accurate assignment for this 
excited state and that electronic spin-orbit interaction mixes in states with $\Lambda = 1$, for example 
a higher-lying ${^1\Pi}_1$ state arising from a $\sigma^1_{6s_{\text{Ta}}}$, $\pi^1_{5d_{\text{Ta}}}$
electronic configuration.

\subsubsection{${\cal{P,T}}$-odd constants and magnetic hyperfine interaction}

We proceed with a detailed discussion of the results presented in Table \ref{TAB:TAN:PROPER}.

\begin{table}[h]

\caption{\label{TAB:TAN:PROPER}
         Vertical excitation energy, molecular electric dipole moment, EDM effective electric field, 
         magnetic hyperfine interaction
         constant, scalar-pseudoscalar electron nucleon interaction constant, and nuclear magnetic quadrupole
         interaction constant at an internuclear distance of $R = 3.1806$ a$_0$ for $\Omega = 1$
        }

\begin{center}
\begin{tabular}{l|ccccccc}
        CI Model     & $T_v$ [\cm] & $D$ [Debye] & $E_{\text{eff}} \left[\frac{\rm GV}{\rm cm}\right]$ & $A_{||}$ [MHz] 
                     & $W_{P,T}$ [kHz] & $W_{M}$ [$\frac{10^{33} {\text{Hz}}}{e\, {\text{cm}}^2}$] \\ \hline
 MR$_{12}$-CISD($10$)       &  $2502$  & $-5.11$ &  $30.1$  &  $-3118$ &  $27.4$  & $0.633$  \\
 MR$_{12}$-CISDT($10$)      &  $2688$  & $-5.13$ &  $29.7$  &  $-3092$ &  $27.1$  & $0.626$  \\
 MR$^{+T}_{12}$-CISD($10$)  &  $2673$  & $-4.81$ &  $31.4$  &  $-3067$ &  $28.7$  & $0.645$  \\
 MR$_{12}'$-CISD($18$)      &  $1972$  & $-5.18$ &  $33.3$  &  $-3147$ &  $30.3$  & $0.714$  \\
 MR$_{12}$-CISD($18$)       &  $2810$  & $-5.46$ &  $33.6$  &  $-3059$ &  $30.5$  & $0.718$  \\
 MR$_{12}$-CISDT($18$)      &  {\it{}} & $-5.49$ &  $33.4$  &  $-3074$ &  $30.3$  & $0.716$  \\
 MR$^{'+T}_{12}$-CISD($18$) &  {\it{}} & $-4.83$ &  $35.1$  &  $-3025$ &  $31.9$  & $0.737$  \\
 MR$^{+T}_{12}$-CISD($18$)  &  {\it{}} & $-4.97$ &  $35.2$  &  $-2917$ &  $32.0$  & $0.739$  \\
 MR$_{12}'$-CISD($26$)      &  $1958$  & $-5.18$ &  $34.2$  &  $-3237$ &  $31.1$  & $0.732$  \\ \hline
\end{tabular}

\end{center}
\end{table}

For all three ${\cal{P,T}}$-odd constants the general trend is observable that their value increases with the
number of explicitly correlated electrons. Comparing $10$-electron with $26$-electron models, the increase
amounts to roughly $14$\% for $E_{\text{eff}}$, $12$\% for $W_{P,T}$, and $14$\% for $W_{M}$, respectively.
The effect of Double excitations from the $5s,5p_{\text{Ta}}$ shells is seen to be negligible for these
three properties, indicating that electron correlations of valence and core-valence type are of predominant 
importance. Since the full 26-electron model is computationally too expensive, the final results will be based
on extensive 18-electron results with an additive correction for $4s,4p_{\text{Ta}}$ core-valence correlation.

Triple excitations into the set of virtual spinors lead to a slight decrease of the ${\cal{P,T}}$-odd constants.
However, this correction is a function of the number of correlated electrons and becomes exceedingly small
for 18-electron models, not surpassing $1$\% in any of the ${\cal{P,T}}$-odd properties. Due to the relatively
large active spinor space, the model MR$_{12}$-CISDT($18$) contains a significant subset of Quadruple and
Quintuple excitations with respect to the simple Fermi vacuum for the $\Omega=0$ electronic ground state.
Therefore, the neglected set of Quadruple and Quintuple excitations comes, in a perturbative sense, with 
large energy denominators. Consequently, such higher excitations (and those beyond) are expected to lead to 
negligble corrections. We include the calculated external Triples corrections in our final property values. 

In subsection \ref{SUBSUBSEC:SPECPROP} we elucidated how a subset of Triple, Quadruple, and Quintuple excitations
involving additional holes in the spinors of $2p_{\text{N}}$/$d_{\text{Ta}}$ character are included in the 
wavefunction expansion. These atomic contributions to these spinors are strongly mixed by the molecular field, and 
the additional higher excitations prove to give non-negligible contributions to the ${\cal{P,T}}$-odd properties.
These contributions are fairly consistent when comparing 10-electron and 18-electron models. $E_{\text{eff}}$
and $W_{P,T}$ are increased by about $5$\%, the MQM interaction constant by about $3$\%, respectively. 
The magnetic hyperfine interaction constant decreases, on the absolute, by about the same percentage.

The magnetic hyperfine coupling constant displays an overall much less systematic behavior than the ${\cal{P,T}}$-odd 
constants. On the absolute, core-valence correlations from the $5s,5p_{\text{Ta}}$ shells leads to a slight 
increase of $A_{||}$ which, however, is overcompensated when including Double excitations from these shells, 
leading to an overall decrease. Core-valence correlations from the $(4s,4p)_{\text{Ta}}$ shells again give a 
slight increase of $A_{||}$. In total, $A_{||}$ varies by only about $4$\% between the models MR$_{12}$-CISD($10$) 
and MR$_{12}^{'}$-CISD($26$). The effect of including Triple excitations into the virtual spinors leads to a
change of less than $1$\%, which again is unsystematic. 

The effect of using $\Delta$-spinors as compared to the state-averaged spinors used throughout in the present
study has been investigated with the MR$_{12}$-CISD($18$) wavefunction expansion. $A_{||}$ is decreased by
$2.7$\% (or $82$ MHz) on the absolute and $W_M$ is decreased by $1.9$\% 
(or $0.01$ $\frac{10^{33} {\text{Hz}}}{e\, {\text{cm}}^2}$), respectively. All other properties 
($D$, $E_{\text{eff}}$, $W_{P,T}$) change by much less than $1$\%.
These corrections, too, will be accounted for in the final estimate.

Table \ref{TAB:TAN:PROPERCORR} presents our final property values for TaN, starting out from the results for model
MR$^{+T}_{12}$-CISD($18$) and including additive corrections accounting for external Triple excitations, 
$(4s,4p)_{\text{Ta}}$ core-valence correlation, and the change to $\Delta$-spinors. The most substantial correction is 

\begin{table}[h]

\caption{\label{TAB:TAN:PROPERCORR}
         Final property values, calculated from values in Table \ref{TAB:TAN:PROPER} including corrections
        }

\vspace*{0.5cm}
\hspace*{-1.7cm}
\begin{tabular}{lrrrrr|l}
 &   $D$ [Debye] & $E_{\text{eff}} \left[\frac{\rm GV}{\rm cm}\right]$ & $A_{||}$ [MHz]
                     & $W_{P,T}$ [kHz] & $W_{M}$ [$\frac{10^{33} {\text{Hz}}}{e\, {\text{cm}}^2}$] \\ \hline
 & $-4.97$ &  $35.2$  &  $-2917$ &  $32.0$  & $0.739$   &   base value from MR$^{+T}_{12}$-CISD($18$)  \\ \hline
 & $+0.04$ &  $-0.2$  &  $-15  $ &  $-0.2$  & $-0.002$  &   external Triples correction \\
 & $+0.02$ &  $+0.1$  &  $+68  $ &  $+0.2$  & $-0.013$  &   correction for $\Delta$ spinors \\
 & $+0.0$  &  $+0.9$  &  $-90  $ &  $+0.8$  & $+0.018$  &   $4s,4p$ core-valence correlation \\ \hline
 & $-4.91$ &  $36.0$  &  $-2954$ &  $32.8$  & $0.742$   & {\bf{Final value}} \\ \hline\hline
 Skripnikov {\it{et al.}}\footnote{Reference \cite{Skripnikov_TaN_PRA2015}} & $4.74$ & $34.9$ & $-3132$ & $31$ &  $1.08$ &  \\
 Flambaum {\it{et al.}}\footnote{Reference \cite{Flambaum_DeMille_Kozlov2014}} &  &  &  &  &  $\approx 1$ & estimate
\end{tabular}
\end{table}

to $E_{\text{eff}}$ and $W_{P,T}$ due to CV-correlation, leading to an increase by $2.5$\%, respectively. This contribution 
is also non-negligible for the magnetic hyperfine constant, but it is here largely counteracted by the $\Delta$-spinor
correction, which is not the case for $E_{\text{eff}}$ or $W_{P,T}$. The MQM interaction constant is left largely
unchanged by these corrections, in total.

The comparison with literature values (Skripnikov {\it{el al.}} \cite{Skripnikov_TaN_PRA2015}) reveals a good
agreement for $E_{\text{eff}}$ and $W_{P,T}$, the values displaying differences of about $3$\% and $5$\%, respectively.
A large part of this difference seems to be due to the $(4s,4p)_{\text{Ta}}$ core-valence correction which has been taken 
into account in the present study.
The difference for $A_{||}$ is roughly $5.5$\%, whereas a sizeable deviation is observed for the MQM interaction
constant, amounting to $\sim30$\%. Since the agreement between other ${\cal{P,T}}$-odd constants is good, there is
no obvious reason - such as a substantial difference in the employed molecular wavefunctions - for the observed
deviation.

\section{Conclusion}
\label{SEC:SUMM}
We summarize and conclude from our most important findings in the present study.
TaN provides an effective electric field of about $36 \left[\frac{\rm GV}{\rm cm}\right]$, which is roughly
as strong as the field obtained recently for ThF$^+$ \cite{ThF+_NJP_2015,Skripnikov_ThF+_PRA2015}, significantly
stronger than the field in YbF \cite{Mosyagin_YbF1998}, but about half the field obtained for ThO
\cite{Fleig2014,Skripnikov_ThO_JCP2015}. The two latter molecules have been used to obtain stronger
constraints on possible sources of EDMs \cite{ACME_ThO_eEDM_science2014,hudson_hinds_YbF2011}.
The nuclear MQM interaction constant obtained in the present work is about $0.74$ 
[$\frac{10^{33} {\text{Hz}}}{e\, {\text{cm}}^2}$], smaller than the estimate by Flambaum {\it{et al.}}
\cite{Flambaum_DeMille_Kozlov2014} but still large enough to hold promise for investigating BSM ${\cal{P,T}}$-odd
hadron physics. To this end, all of the experimental advantages of using molecules in a low-lying {$^3\Delta$} 
state may be exploited. The very large molecule-frame dipole moment of $-4.91$ [Debye] we obtain allows
for efficient polarization of the molecule in a weak external electric field.

\begin{acknowledgments}
TF and MKN thank the {\it{Agence Nationale de la Recherche}} (ANR) through grant no.
ANR-BS04-13-0010-01, project ``EDMeDM'', for financial support.
We wish to thank David DeMille (Yale) for helpful discussions.

\end{acknowledgments}

\clearpage



\bibliographystyle{unsrt}
\newcommand{\Aa}[0]{Aa}

\end{document}